# Diagnosis and Prediction of the 2015 Chinese Stock Market Bubble

Min Shu[1, 2, *], Wei Zhu[1, 2]

[1] Department of Applied Mathematics & Statistics, Stony Brook University, Stony Brook, NY, USA
[2] Center of Excellence in Wireless & Information Technology, Stony Brook University, Stony Brook, NY, USA

## Abstract

In this study, we perform a novel analysis of the 2015 financial bubble in the Chinese stock market by calibrating the Log Periodic Power Law Singularity (LPPLS) model to two important Chinese stock indices, SSEC and SZSC, from early 2014 to June 2015. The back tests of the 2015 Chinese stock market bubbles indicates that the LPPLS model can readily detect the bubble behavior of the faster-than-exponential increase corrected by the accelerating logarithm-periodic oscillations in the 2015 Chinese Stock market. The existence of log-periodicity is detected by applying the Lomb spectral analysis on the detrended residuals. The Ornstein-Uhlenbeck property and the stationarity of the LPPLS fitting residuals are confirmed by the two Unit-root tests (Philips-Perron test and Dickery-Fuller test). According to our analysis, the actual critical day $t_c$ can be well predicted by the LPPLS model as far back as two months before the actual bubble crash. Compared to the traditional optimization method used in the LPPLS model, we find the covariance matrix adaptation evolution strategy (CMA-ES) to have a significantly lower computation cost, and thus recommend this as a better alternative algorithm for LPPLS model fit. Furthermore, in the LPPLS fitting with expanding windows, the gap ($t_c$ -$t_2$) shows a significant decrease when the end day $t_2$ approaches the actual bubble crash time. The change rate of the gap ($t_c$ -$t_2$) may be used as an additional indicator besides the key indicator $t_c$ to improve the prediction of bubble burst.

*Corresponding author at: Department of Applied Mathematics & Statistics, Physics A149, Stony Brook University, Stony Brook, NY 11794, USA.
*E-mail address:* min.shu@stonybrook.edu (M. Shu), wei.zhu@stonybrook.edu (W. Zhu)

## 1. Introduction

Financial bubbles and crashes are traumatic events in the modern society often with far reaching consequences. In the past 30 years there are approximately 100 financial crises worldwide (Stiglitz, 2014). Painful memories have taught us that it is vital to identify bubbles in time, limit their sizes, all in the hope to minimize the damages from the bursts. However, due to the complexity of the local and global economies and the increasing correlations among world financial markets, it is arduous to characterize, forecast and possibly avoid bubbles in advance.

Financial bubbles are generally defined as the accelerating ascent of an asset price above the fundamental value of the asset. However, the fundamental value of an asset is generally not sufficiently constrained rendering it taxing to distinguish between an exponentially growing fundamental price and an exponentially growing bubble price. Therefore, the problem of unambiguous identification of the presence of a bubble remains unresolved in standard econometric and financial economic approaches (Gurkaynak, 2008; Lux and Sornette, 2002). In order to detect the presence of a bubble, a more precise definition of bubbles is required to overcome the above issues.

The causes of bubbles have been widely investigated and many theories have been developed to explain the potential causes of bubbles. Financial bubbles can be generated by the irrationality of investors. Galbraith (2009) stated that the stock market can be raised by "the vested interest in euphoria that leads men and women, individuals and institutions to believe that all will be better, that they are meant to be richer and to dismiss as intellectually deficient what is in conflict with that conviction". Shiller (2015) believed that the price of stock was driven high by irrational euphoria among individual investors who were catered by the pseudo-news from an emphatic media. Even if there are no irrational investors, the bubbles of stock market in recent theories can be generated due to (1) heterogeneous beliefs of investors together with short-time constraints, (2) positive feedback trading by noise traders, and, (3) synchronization failures among rational traders.

To improve the traditional definition of a bubble, the Log Periodic Power Law Singularity (LPPLS) model (Johansen & Sornette, 1999a, 2000; Sornette, 2009) has been developed to define the bubble in an alternative way. Instead of describing bubbles by exponential prices, the bubbles are characterized by a faster-than-exponential (or super-exponential) growth of price leading to unsustainable growth ending with a finite crash-time $t_c$ in the LPPLS model. The reason for the super-explosive growth of price of a bubble is that positive feedback in the valuation of assets created by imitation and herding behavior of noise traders and of boundedly rational agent results in price processes that exhibit a finite-time singularity at some future time (Yan, 2011). Due to the tension and competition between the value investors and the noise traders, the market price of an asset is deviated around the faster-than-exponential growth in the form of oscillations that are periodic in the logarithm of the time to $t_c$. The LPPLS model provided a quantitative framework to detect financial bubbles by analyzing the price time series of an asset. However, the LPPLS model is suitable only for endogenous crashes that constitute about two-thirds of crashes. Endogenous crashes are preceded by bubbles generated through positive-feedback mechanisms dominated by the imitation and herding of the noise traders. Over the past decade, the LPPLS model has been widely used to detect bubbles and crashes in advance in various markets, such as the 2006-2008 oil bubble (Sornette et al., 2009), the Chinese stock market bubbles in 2005–2007 and 2008–2009 (Jiang et al., 2010), the real estate market bubble in Las Vegas (Zhou and Sornette, 2008), the 2000-2003 real estate bubble in the UK (Zhou and Sornette, 2003), the USA real estate bubble (Zhou and Sornette, 2006) and the S&P 500 index anti-bubble in 2000-2003 (Johansen and Sornette, 1999).

The LPPLS model has seen much interest and research in the recent years. Yan et al. (2012) tailored the LPPLS model of rational expectation bubbles to model the negative bubbles in order to detect the rebounds of financial markets. Brée et al. (2013) demonstrated that the LPPLS functions are intrinsically very hard to fit to time series by taking into account the sloppiness. Filimonov and Sornette (2013) suggested a simple transformation of the formulation of the LPPLS model by reducing the number of nonlinear parameters. Lin et al. (2014) proposed a self-consistent model for explosive financial bubbles which combines a mean-reverting volatility process and a stochastic conditional return. Sornette et al. (2015) gauged the performance of the real-time prediction and post-mortem analysis of Shanghai stock market bubble regime that started to burst in June 2015. Zhang et al. (2016) developed novel tests for the early causal diagnostic of positive and negative bubbles and the detection of the end signals of bubbles using the monthly S&P 500 data from August 1791 to August 2014. Li (2017) used the Shanghai Shenzhen CSI 300 index to analyze the critical dates of three historical Chinese stock market bubbles, and suggested that the LPPLS is available to predict the bubble crashes and the forecast gap is an alternative indicator of the bubble process sustainability. Demos and Sornette (2017) performed systematic tests to determine the precision and reliability of the beginning and end time of a bubble, and concluded that it is much better to constrain the beginning of bubbles than their end. Filimonov et al. (2017) presented the modified profile likelihood inference method for the calibration of LPPLS model and the interval estimation of the critical time.

In the past three decades, the Chinese economy has seen a tremendous growth accompanied by a roller coaster ride of the Chinese stock markets, with three large bubbles bursting respectively from May 2005 to October 2007, from November 2008 to August 2009 and from mid-2014 to June 2015 (Sornette et al., 2015). In mainland China, the organized stock market is composed of two stock exchanges: Shanghai stock exchange (SHSE) and Shenzhen stock exchange (SZSE). The Shanghai stock exchange composite index (SSEC) and the Shenzhen stock exchange component index (SZSC) are the most important indices for A-shares in SHSE and SZSE. Due to profiting from the easy access to credit to invest in the stock markets, about 7% of China’s population has been active in stock market. In nearly five and a half months from December 31, 2014 to June 12, 2015, the SSEC and SZSC indices soared by 60% and 122%, respectively, while the Chinese overall economy was cooling significantly at the time. The 2015 Chinese Stock Market bubble can be seen as a result of a strong leverage that the realities of economic activity is disconnected from the corporate earnings.

The 2015 Chinese Stock Market bubble crashed on June 12, 2015. The SSEC index has suffered more than 43% drop from the peak on June 12, 2015 to the bottom on August 26, 2015, and SZSC index has lost 45% over the same period. Figure 1 shows the time evolution of the price trajectories of the SSEC index and the SZSC index in the 2015 Chinese Stock Market bubble.

The paper is organized as follows. Section 2 presents the technical descriptions of all the methods used in this study, including the LPPLS model, LPPLS calibration, Lomb periodogram analysis, unit root tests and change-of-regime statistics. The empirical analysis of the 2015 Chinese Stock Market bubble are conducted in Section 3. Section 4 concludes.

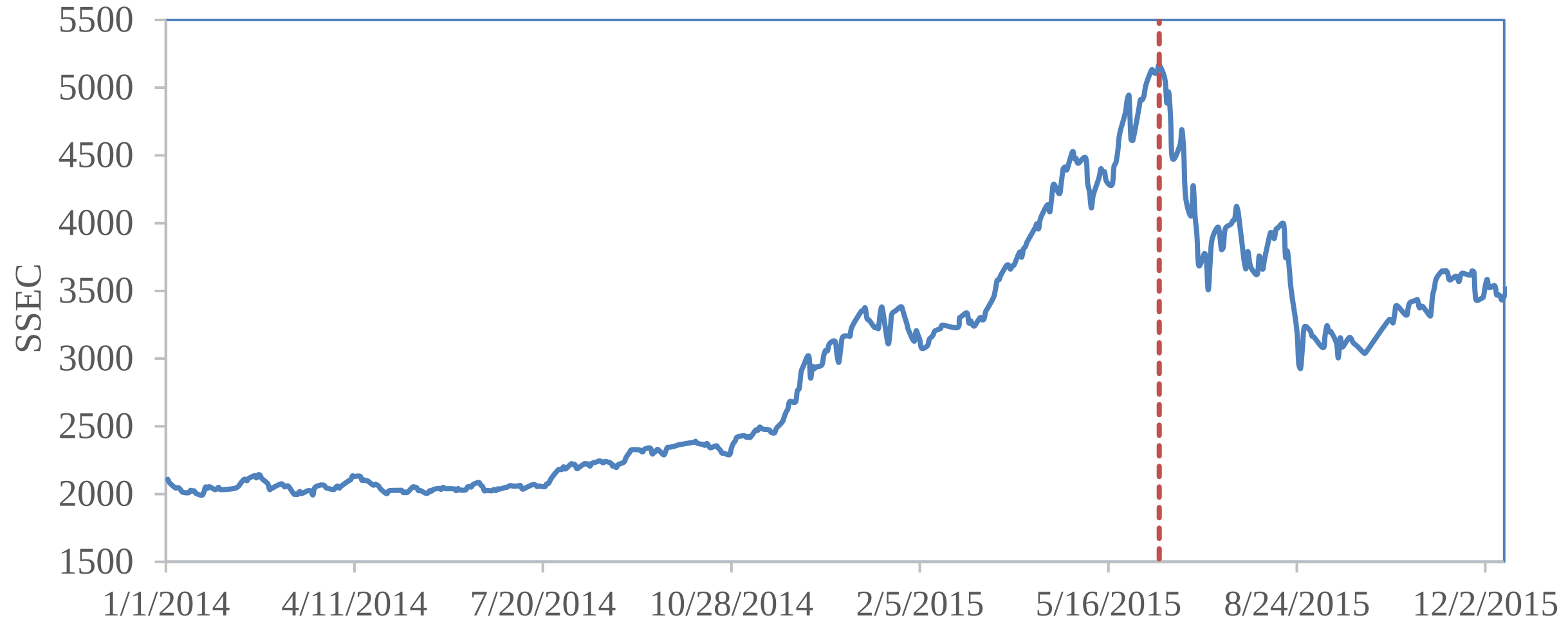

(a) SSEC index

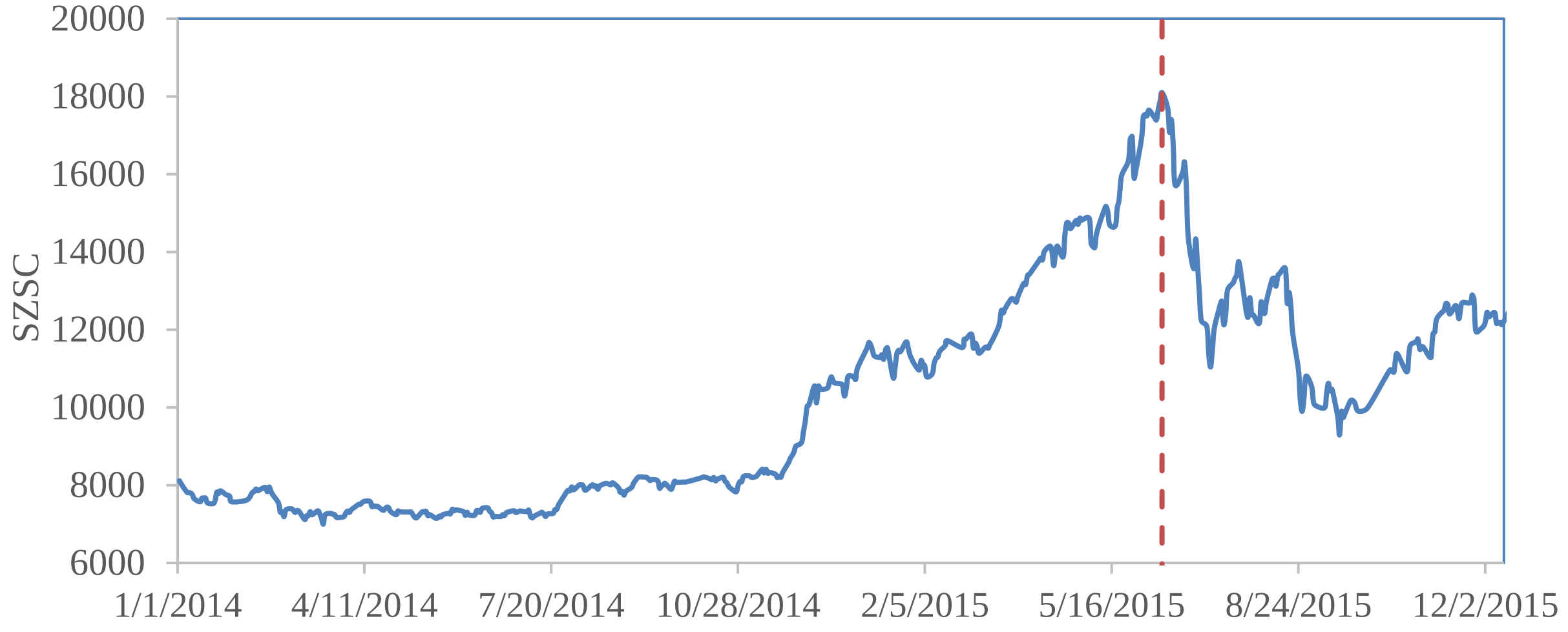

(b) SZSC index

Figure 1. Evolution of the price trajectories of the SSEC index and the SZSC index before and after the 2015 Chinese stock market crash.

## 2. Methodology

The main method for predicting the critical time $t_c$ when the bubble will end in either a crash or change of regime is by fitting the observed financial index price time series to a log-periodic power law singularity (LPPLS) model (Sornette, 2003). In contrast to traditional optimization algorithms, we recommend the covariance matrix adaptation evolution strategy (CMA-ES) algorithm as a better LPPLS model fitting procedure, as described below.

*2.1 The Log-Periodic Power Law Singularity (LPPLS) Model*

The LPPLS model is an extension of the rational expectation bubble model (Blanchard and Watson, 1982). A financial bubble is modeled as a process of super-exponential power law growth punctuated by short-lived corrections organized according to the symmetry of discrete scale invariance (Sornette, 1998). The LPPLS model combines (i) the economic theory of rational expectation bubbles, (ii) behavioral finance on imitation and herding of traders, and (iii) the mathematical and statistical physics of bifurcations and phase transitions (Yan, 2011). The LPPLS model takes into account the faster-than-exponential growth in asset prices as well as the accelerating logarithm-periodic oscillations to detect the bubbles. It is assumed that the observed price trajectory of a given asset decouples from its intrinsic fundamental value in a bubble regime (Sornette, 2003). For a given fundamental value, the Johansen-Leoit-Sornette (JLS) model (Johansen et al., 2000) assumes that the logarithm of the observed asset price $p(t)$ can be expressed as:

$$\frac{dp}{p} = \mu(t)dt + \sigma(t)dW - kdj \qquad (1)$$

where $\mu(t)$ is the expected return, $\sigma(t)$ is the volatility, $dW$ is the infinitesimal increment of a standard Wiener process, $k$ is the loss amplitude of a possible crash, and $dj$ represents a discontinuous jump with the value of 0 before the crash and 1 after the crash. The dynamics of the jumps is governed by a crash hazard rate $h(t)$, which is the crash probability at a specified time $t$. Since $h(t)dt$ is the probability that the crash occurs between $t$ and $t + dt$ conditional on the fact that it has not yet happened, the expectation of $dj$ can be determined as: $E[dj] = h(t)\mathrm{d}t$.

In the LPPLS model, it is assumed that two groups of agents are present in a market: one group of traders with rational expectations and the other group of noise traders who may destabilize the asset price due to imitation and herding behavior. According to the LPPLS model, the aggregate effect of noise traders can be quantified by the following dynamics of the hazard (Johansen et al., 2000):

$$h(t) = \alpha(t_c - t)^{m-1}(1 + \beta cos(\omega\, ln(t_c - t) - \phi)) \qquad (2)$$

where $\alpha, \beta, m, \omega,$ and $\varphi'$ are the parameters. The power law behavior $(t_c - t)^{m-1}$ embodies the mechanism of positive feedback which results in the formation of bubbles. The log periodic function $\cos(\omega \ln(t_c - t) - \phi)$ accounts for the existence of a possible hierarchical cascade of panic acceleration punctuating the growth of the bubble, resulting either from a preexisting hierarchy in noise trader sizes (Sornette and Johansen, 1997) and/or from the interplay between market price impact inertia and nonlinear fundamental value investing (Ide and Sornette, 2002).

The non-arbitrage condition expresses that the unconditional expectation $E[dp]$ of the price increment should be 0, resulting in:

$$\mu(t) \equiv E[\frac{dp/dt}{p}]_{\text{no crash}} = kh(t) \quad (3)$$

Solving Equation (1) by substituting Equation (2) and Equation (3) and under the condition that no crash has yet occurred leads to the simple mathematical formulation of the LPPLS for the expected value of a log-price (Sornette, 2003):

$$\text{LPPLS}(t) \equiv \ln E\,[p(t)] = A + B(t_c - t)^m + C(t_c - t)^m \cos[\,\omega \ln(\,t_c - t) - \phi] \quad (4)$$

where $B = -k\alpha/m$ and $C = -k\alpha\beta/\sqrt{m^2 + \omega^2}$. The critical time $t_c$ corresponding to the theoretical termination of a financial bubble indicates the change to another regime, which could be a large crash or a change of the average growth rate. The bubble regimes are in general characterized by $0 < m < 1$ and $B < 1$. The first condition of $m > 0$ ensures that the price remains finite at the critical time $t_c$, while $m < 1$ indicates that a singularity exists. The two considerations ensure that the price is indeed growing super-exponentially as time goes towards $t_c$.

*2.2 LPPLS fitting technique*

The LPPLS model in its original form in Equation (4) presents a function with 3 linear parameters ($A$, $B$, and $C$) and 4 nonlinear parameters ($t_c, m, \omega$ $and$ $\phi$) which should be estimated by fitting the function to the observed price time series within a time window. Due to the relatively large number of parameters and the strong nonlinear structure of the model, it is a non-trivial task to calibrate the LPPLS model and multiple local minima could be obtained leading to the local optimization algorithm getting trapped. Most of the fitting procedures subordinate the 3 linear parameters to the 4 nonlinear parameters and the search space is reduced to the 4-dimensional parameter space. However, it is still difficult to calibrate the LPPLS model since the search space with 4 nonlinear parameters has a very quasi-periodic structure with multiple minima. Some meta-heuristic methods such as taboo search (Cvijovic & Klinowski, 1995) or genetic algorithm (Jacobsson, 2009) have to be used to determine the global minimum. Even so, the correct solution may not be discovered. In addition, the issue is not satisfactorily solved on how to deal with the existence of many possible competing degenerate solutions.

Filimonov and Sornette (2013) proposed a fundamental revision of the formulation of the LPPLS model to transform it from a function of 3 linear and 4 nonlinear parameters into a representation with 4 linear and 3 nonlinear parameters. This reformulation of the LPPLS model decreased the number of nonlinear parameters and removed the interdependence between the angular log-frequency $\omega$ and the phase $\phi$. It can be described as:

$$\text{LPPLS}(t) = \ln E[p(t)] = A + B(t_c - t)^m + C_1(t_c - t)^m \cos(\omega \ln(t_c - t)) + C_2(t_c - t)^m \sin(\omega \ln(t_c - t)) \quad (5)$$

where $C_1 = C\cos\phi$ and $C_2 = C\sin\phi$. The reformed LPPLS model has now only 3 nonlinear parameters ($t_c, m, \omega$) and 4 linear parameters ($A, B, C_1, C_2$), and the phase $\phi$ is contained by $C_1$ $and$ $C_2$. The cost function in the least-squares method can be described as:

$$F(t_c, m, \omega, A, B, C_1, C_2) = \sum_{i=1}^{N} [\ln p(\tau_i) - A - B(t_c - \tau_i)^m - C_1 (t_c - \tau_i)^m \cos(\omega \ln(t_c - \tau_i)) - C_2 (t_c - \tau_i)^m \sin(\omega \ln(t_c - \tau_i))]^2 \quad (6)$$

where $\tau_1 = t_1$ and $\tau_N = t_2$. Slaving the 4 linear parameters $A, B,\ C_1\ and\ C_2$ to the 3 nonlinear parameters $t_c, m,\ \omega$, the nonlinear optimization problem is:

$$\{\hat{t}_c, \hat{m}, \hat{\omega}\} = arg \min_{t_c, m, \omega} F_1(t_c, m, \omega)$$

where the $F_1(t_c, m, \omega)$ is given by $F_1(t_c, m, \omega) = \min_{A, B,\ C_1, C_2} F_1(t_c, m, \omega, A, B,\ C_1, C_2)$. The linear parameters can be solved by:

$$\begin{pmatrix} N & \sum f_i & \sum g_i & \sum h_i \\ \sum f_i & \sum f^2_i & \sum f_i g_i & \sum f_i h_i \\ \sum g_i & \sum f_i g_i & \sum g^2_i & \sum h_i g_i \\ \sum h_i & \sum f_i h_i & \sum g_i h_i & \sum h^2_i \end{pmatrix} \begin{pmatrix} \hat{A} \\ \hat{B} \\ \hat{C}_1 \\ \hat{C}_2 \end{pmatrix} = \begin{pmatrix} \sum \ln p_i \\ \sum f_i \ln p_i \\ \sum g_i \ln p_i \\ \sum h_i \ln p_i \end{pmatrix} \quad (7)$$

where $f_i = (t_c - \tau_i)^m$, $g_i = (t_c - \tau_i)^m cos(\omega \ln(t_c - \tau_i))$, and $h_i = (t_c - \tau_i)^m sin(\omega \ln(t_c - \tau_i))$. The reformulation of the LPPLS model decreases the complexity of the fitting procedure while improves its stability tremendously because the modified cost function is characterized by good smooth properties. With the methodology proposed by Filimonov and Sornette (2013), the meta-heuristics are no longer necessary and one can resort solely to rigorous controlled local search algorithms, leading to some dramatic increase in computational efficiency.

In order to minimize the fitting problems and address the sloppiness of the model with respect to some of its parameters (Brée et al., 2013), we use the following filters to condition the solutions:

$$m \in [0.1, 0.9], \omega \in [6, 13], t_c \in [t_2, t_2 + (t_2 - t_1)/3],\ m|B|/(\omega\sqrt{C_1^2 + C_2^2}) \geq 1, \\ (\omega/\pi)\ln[(t_c - t_1)/(t_c - t_2)] \geq 2.5 \quad (8)$$

These filters to bound the search space derived from the empirical evidence gathered in investigations of previous bubbles are the stylized features of LPPLS model. The more stringent constraint $m \in [0.1, 0.9]$ improves the power of discriminating bubbles to avoid the select parameters being too close to the bounds (Demos and Sornette, 2017). The condition $\omega \in [6, 13]$ constrains the log-periodic oscillations to be neither too fast to fit the random component of the data, nor too slow to provide a contribution to the trend (Huang et al., 2000). The condition $t_c \in [t_2, t_2 + (t_2 - t_1)/3]$ ensures that the predicted critical time $t_c$ should be after the end $t_2$ of the fitted time series, and the upper bound of $t_c$ should not be too far away from the end of the time series since the predictive capacity degrades far beyond $t_2$ (Jiang et al., 2010). The damping parameter $m|B|/(\omega\sqrt{{C_1}^2 + {C_2}^2} \geq 1$ is due to the condition that the crash hazard rate $h(t)$ is non-negative by definition (Bothmer and Meister, 2003). The condition for the number of oscillations (half-periods) of the log-periodic component $(\omega/\pi)\ln[(t_c - t_1)/(t_c - t_2)] \geq 2.5$ is implemented to distinguish a genuine log-periodic signal from one that could be generated by noise (Huang et al., 2000).

*2.3 The covariance matrix adaptation evolution strategy*

In this study, the covariance matrix adaptation evolution strategy (CMA-ES) is applied to search the best estimation of the three nonlinear parameters $(t_c, m, \omega)$ to minimize the residuals (the sum of the squares of the differences) between the fitted LPPLS model and the observed price time series. The CMA-ES, proposed by Hansen and colleagues (Hansen et al., 1995) and further developed by Hansen and colleagues (Auger and Hansen, 2005; Hansen et al., 2003; Hansen and Ostermeier, 2001), is currently the most widely used evolutionary algorithm. The CMA-ES is rated among the most successful evolutionary algorithms for real-valued single-objective optimization is typically applied to difficult nonlinear non-convex black-box optimization problems in continuous domain and search space dimensions between three and a hundred. The main advantages of the CMA-ES lie in its invariance properties, including invariance to order preserving transformations of the objective function value and invariance to angle preserving transformations of the search space if the initial search point is transformed accordingly (Igel et al., 2007).

Compared to the traditional fitting method used in the LPPLS model, such as taboo search, genetic algorithm, Levenberg–Marquardt method, and Nelder-Mead Simplex search method, the CMA-ES may have a lower computation cost and smaller computational error. It is noted that the maximum number of iterations has a significant influence on the computation performance of the CMA-ES. The higher maximum number of iterations may lead to more accuracy of computation, but also result in higher computation cost and longer computation time. In this study, the maximum number of iterations is set to 500. In order to expedite the fitting process, parallel computing is adopted to reduce the computation time remarkably.

*2.4 Stability of fits and probabilistic forecasts*

Since the early deviation of the observed price from its fundamental value is relatively small in the first month and even first years of the bubble, a single beginning date $t_1$ in the fitting time window may be unreliable. To make the prediction more statistically robust, the ensemble of fits with varying window sizes is recommended. In order to test the sensitivity of variable fitting intervals $[t_1, t_2]$, the strategy of fixing one endpoint and varying the other one is adopted. If $t_2$ is fixed, the time window shrinks in terms of $t_1$ moving towards $t_2$ with a step of $dt_1$. If $t_1$ is fixed, the time window expands in terms of $t_2$ moving away from $t_1$ with a step of $dt_2$. Due to the rough nonlinear parameter landscape in the LPPLS model and the stochastic nature of solving multiple dimensional nonlinear optimization problems, a different set of fitting parameters is expected for each implementation of fit process. To investigate an optimal region of solution space, the fitting procedure is repeatedly implemented three times for each window interval. Since the theoretical distribution of $t_c$ is unknown and the sample size may be insufficient for straightforward statistical inference, the bootstrap technique is employed to resample the sample data and perform inference on the sample estimates. Based on sampling many intervals with the bootstrap techniques, the probabilistic forecasts of the critical time $t_c$ can be obtained.

*2.5 Lomb spectral analysis*

In order to detect the logarithm-periodic oscillations in fitting the logarithm of prices to the LPPLS model, the Lomb spectral analysis is used in this study. The Lomb spectral analysis is a spectral analysis designed for irregularly sample data and reaches the same results as the standard Fourier spectral analysis for uniformly spaced data. Given a time series, the Lomb analysis returns a series of frequencies $\omega$ as well as the power at each frequency $P_N(\omega)$. The Lomb frequency $\omega_{Lomb}$ is the frequency with the maximum power. The parametric detrending approach (Sornette and Zhou, 2002) is performed in this study. The series of detrended residual is calculated as:

$$r(t) = (t_c - t)^{-m}(\ln[p(t)] - A - B(t_c - t)^m) \tag{9}$$

As the logarithm-periodic oscillations results from the cosine part in the LPPLS, the Lomb frequency $\omega_{Lomb}$ needs to be compared with the fitted angular frequency $\omega_{\text{fit}}$ in the LPPLS fitting procedure.

*2.6 Ornstein–Uhlenbeck and unit root tests*

According to the study of Lin et al. (2014), the LPPLS fitting residuals can be modeled by a mean-reversal Ornstein-Uhlenbeck (O-U) process if the logarithmic price in the bubble regime is attributed to a deterministic LPPLS component. The test for the O-U property of LPPLS fitting residuals can be translated into an AR(1) test for the corresponding residuals. Hence, the O–U property of fitting residuals can be verified by applying the unit-root tests on the residuals. In this study, both the Phillips-Perron unit-root test and the Dickey-Fuller unit-root test are used to check the O-U property of LPPLS fitting residuals. The rejection of null hypothesis $H_o$ indicates that the residuals are stationary and thus compatible with the O-U process in the residuals. If the null hypothesis of both Phillips-Perron and Dickey-Fuller unit-root tests cannot be rejected, we can be confident that the residual time series has indeed a unit root.

## 3. Empirical analysis

In the following three subsections, we present a novel analysis of the 2015 Chinese Stock Market bubble using the methods described in Section 2.

Figure 1 shows the two-dimensional cross-sections of the cost function in the three nonlinear parameters $\mathrm{t_c}$, $m$, and $\omega$ of the LPPLS formula (5) using the time window from Aug 8, 2014 to April 20, 2015. The best-fit parameters are depicted by the red vertical lines. This figure presents additional information of the nature of the optimization process in the LPPLS fitting procedure. The shape of the cost function allows us to determine the confidence interval for the critical time $t_c$. It is noted that the cost function is convex in the space $m$ and $\omega$, indicating a rather precise determination of $\omega$ may have the significant association with the clear characteristic spells of log-price acceleration.

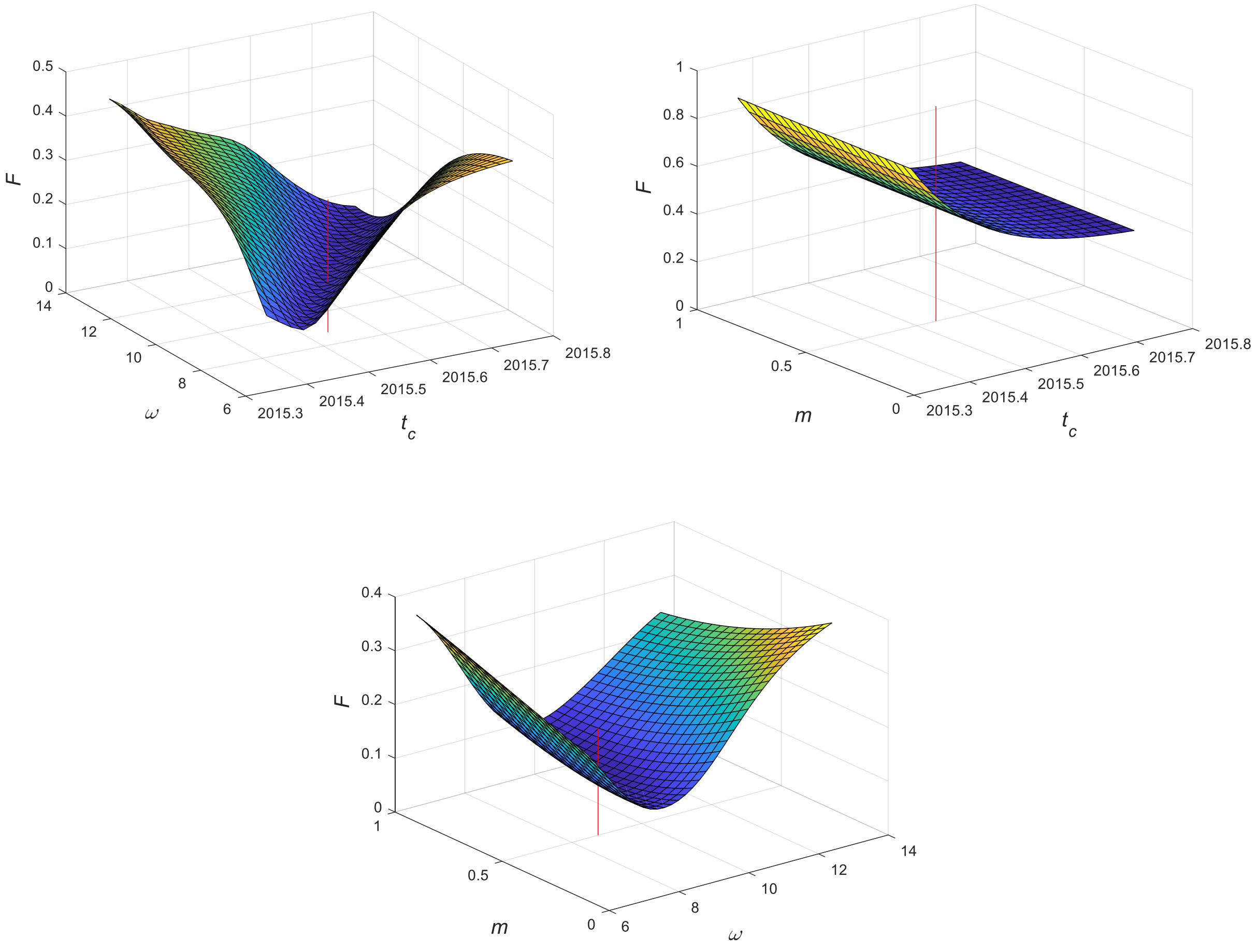

Figure 1. Three cross-sections of the cost function landscape as a function of pairs formed from the three nonlinear parameters $t_c$, $m$, and $\omega$ of the LPPLS formula (5) obtained in the time window [2014/8/8, 2015/4/20].

### *3.1 LPPLS fitting with varying window sizes*

The sensitivity of fit parameters for the two important Chinese stock indices, SSEC and SZSC, is tested by varying the size of the fit intervals. In the expanding windows, the start time $t_1$=November 3, 2014 is fixed with the end date $t_2$ increasing from March 27, 2015 to June 10, 2015 in steps of three trading days. In the shrinking windows, the end time $t_2$=April 20, 2015 is fixed with the start time $t_1$ increasing from January 2, 2014 to January 30, 2015 in steps of three trading days.

In the expanding and shrinking fitting procedures, 18 times in expanding windows and 89 times in shrinking windows are fitted. Based on the LPPLS conditions, 16 (17) and 79 (75) results for SSEC (SZSC) are filtered in expanding and shrinking windows, respectively. Figure 2 (a) illustrates four selected fitting results of the expanding windows for SSEC, and (b) illustrates four chosen fitting examples of the shrinking windows for SSEC. The 20%/80% and 5%/95% quantile range of values of the crash dates $t_c$ are from June 2, 2015 to July 3, 2015 and from May 19, 2015 to July 9, 2015 for the expanding windows. In the figures, the dark shadow box indicates the 20%/80% quantile range of the values of the fitted crash date. For the shrinking windows, the 20%/80% and 5%/95% quantile range of values of the predicted crash dates $t_c$ are from June 5, 2015 to July 13, 2015 and from May 22, 2015 to July 28, 2015, respectively. The observed market peak date for the SSEC is June 12, 2015, which lies in the quantile ranges of the predicted crash dates $t_c$ fitted based on data before the actual stock market crash.

Figure 3 shows the daily trajectory of the logarithmic SZSC index and the sample fits using the LPPLS formula in the expanding (a) and shrinking (b) windows, respectively. The 20%/80% and 5%/95% quantile range of values of the crash dates $t_c$ are from June 9, 2015 to June 24, 2015 and from May 27, 2015 to July 30, 2015 for the expanding windows. For the shrinking windows, the 20%/80% and 5%/95% quantile range of values of the fitted crash dates $t_c$ are from June 5, 2015 to July 16, 2015 and from May 27, 2015 to July 27, 2015, respectively. We see that it is feasible to predict the crash date $t_c$ in the stock market, in advance.

In the LPPLS model, the exponent $m$ captures the mechanism of positive feedback leading to faster-than-exponential price growth. Figure 4 (a) shows the change of $m$ with the increasing end date $t_2$ when the start date $t_1$ is fixed. It can be noted that the exponent $m$ does not show a remarkable feature of change of when varying the expanding windows for SSEC and SZSC. Figure 4 (b) illustrates the variation of the gap between the predicted critical time $t_c$ and the end time of the time interval ($t_c$-$t_2$) when the expanding windows are adopted. As shown in Figure 4 (b), the gap ($t_c$-$t_2$) may be significant decreased when the bubble is closing to crash. This finding is in good agreement with Li (2017). The change of gap ($t_c$-$t_2$) may be used as an additional indicator besides of the key indicator $t_c$ to improve the accuracy of prediction of bubble burst.

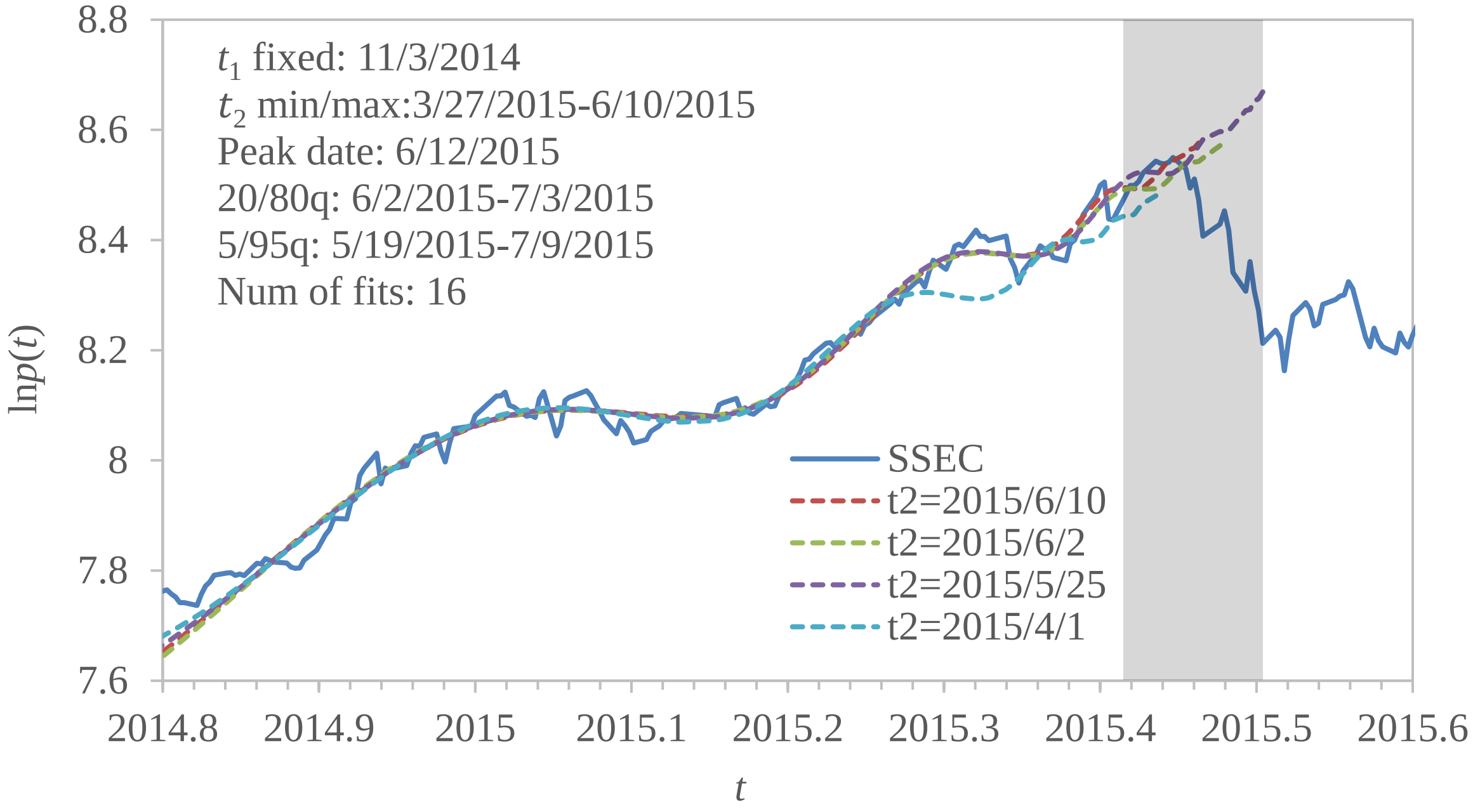

(a) Examples of fitting to the expanding windows with the $t_1$ fixed at November 3, 2014 and varied $t_2$ for SSEC. The four fitting examples are corresponding to $t_2$= 10 June 2015, 2 June 2015, 25 May 2015, and 1 April 2015.

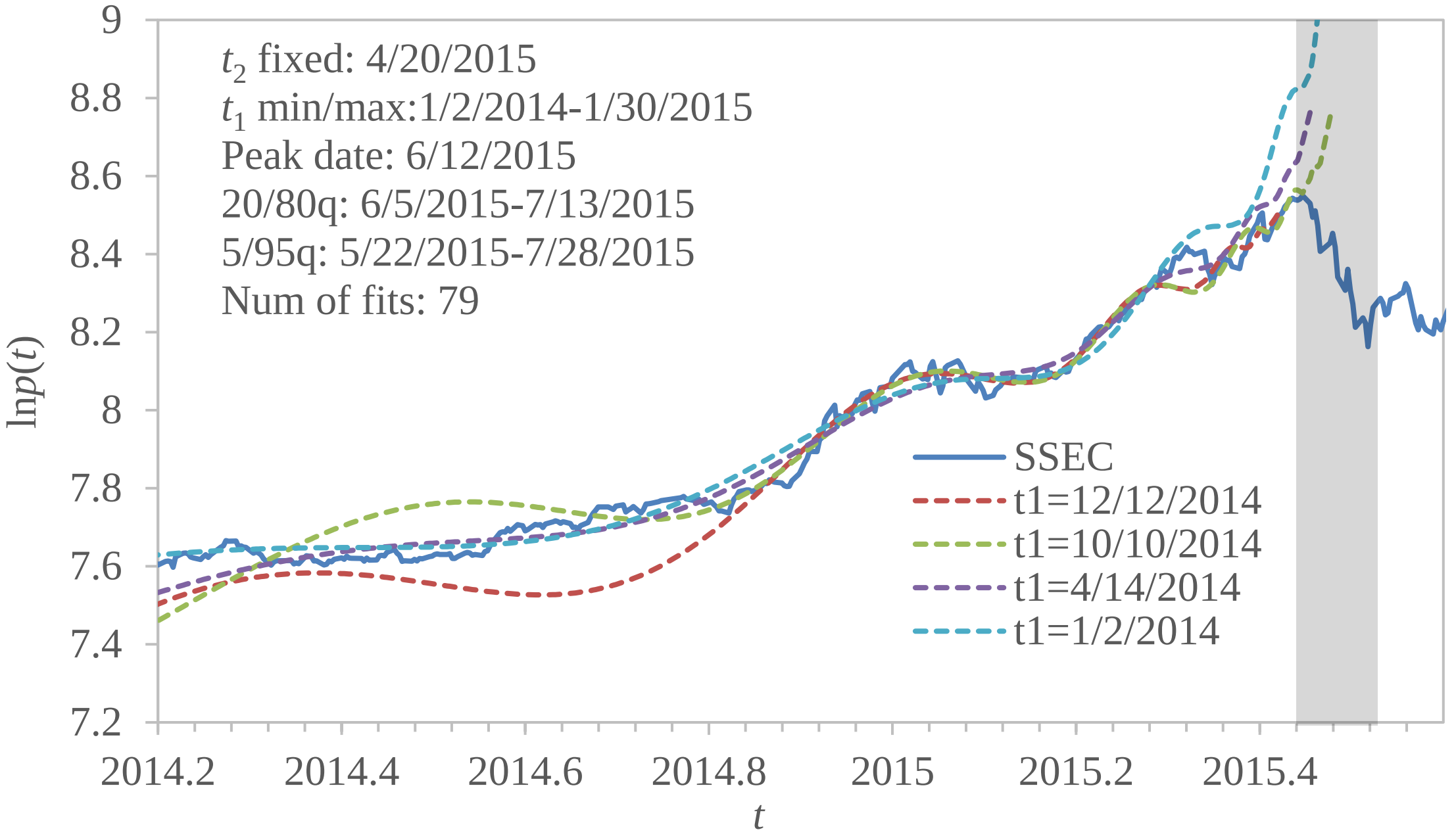

(b) Examples of fitting to the shrinking windows with the $t_2$ fixed at April 20, 2015 and varied $t_2$ for SSEC. The four fitting examples are corresponding to $t_1$= 12 December 2014, 10 October 2015, 14 April 2014, and 2 January 2014.

Figure 2. Daily trajectory of the logarithmic SSEC (a and b) index and the fits using the LPPLS formula. The dark shadow box indicates the 20%/80% quantile range of the fitted crash date.

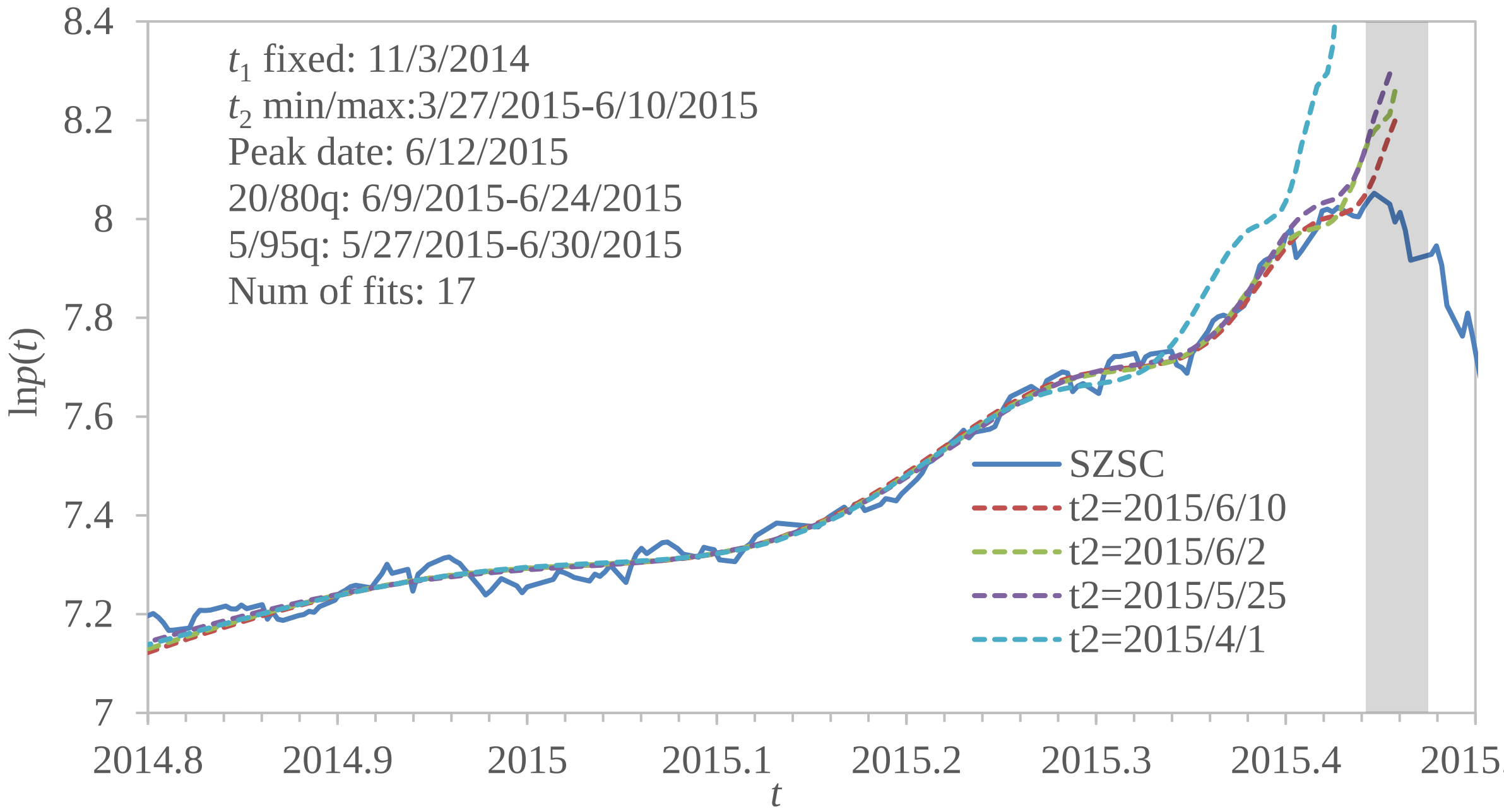

(a) Examples of fitting to the expanding windows with the $t_1$ fixed at November 3, 2014 and varied $t_2$ for SZSC. The four fitting examples are corresponding to $t_2$= 10 June 2015, 2 June 2015, 25 May 2015, and 1 April 2015.

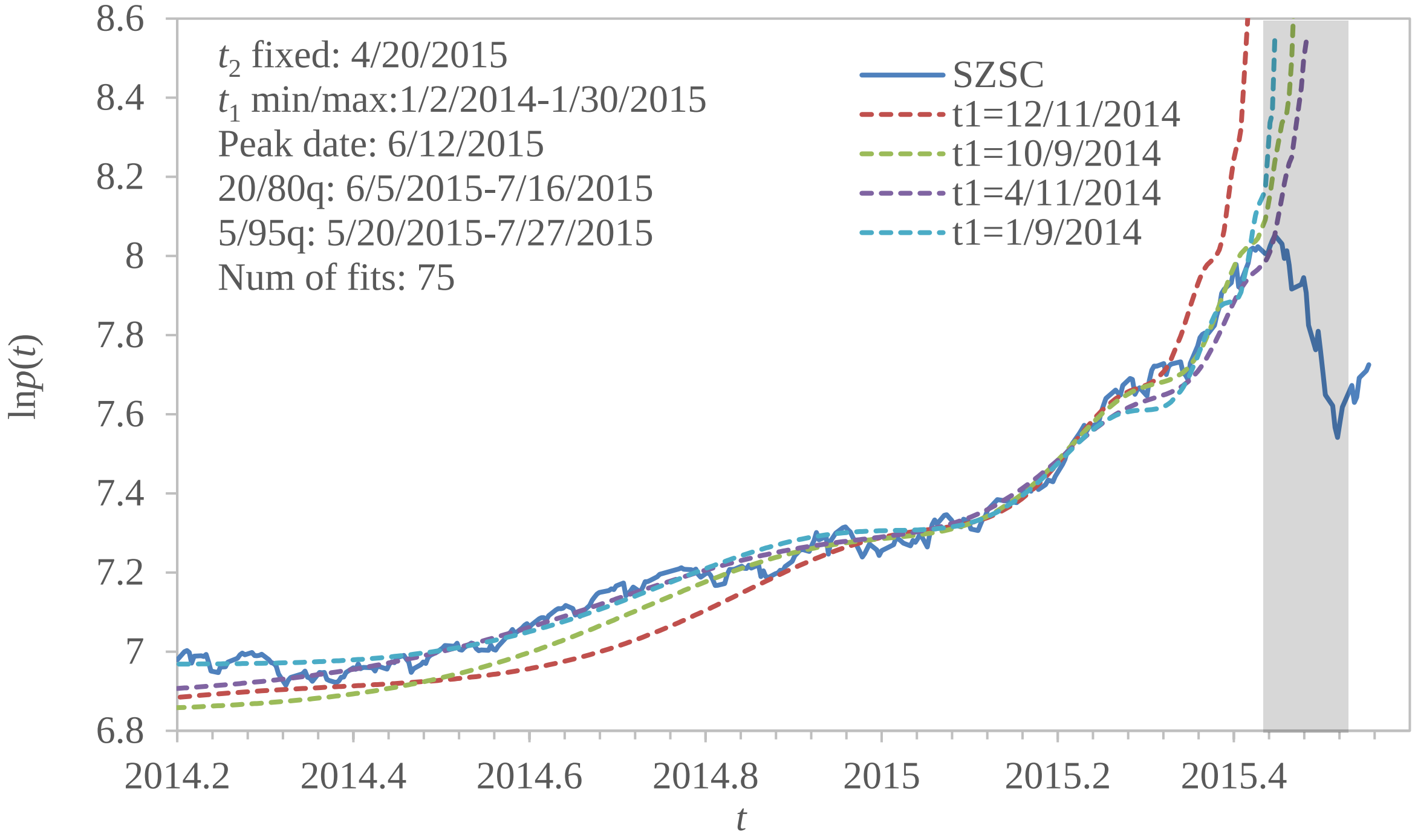

(b) Examples of fitting to the shrinking windows with the $t_2$ fixed at April 20, 2015 and varied $t_2$ for SZSC. The four fitting examples are corresponding to $t_1$= 11 December 2014, 9 October 2015, 11 April 2014, and 9 January 2014.

Figure 3. Daily trajectory of the logarithmic SZSC (a and b) index and the fits using the LPPLS formula. The dark shadow box indicates the 20%/80% quantile range of the fitted crash date.

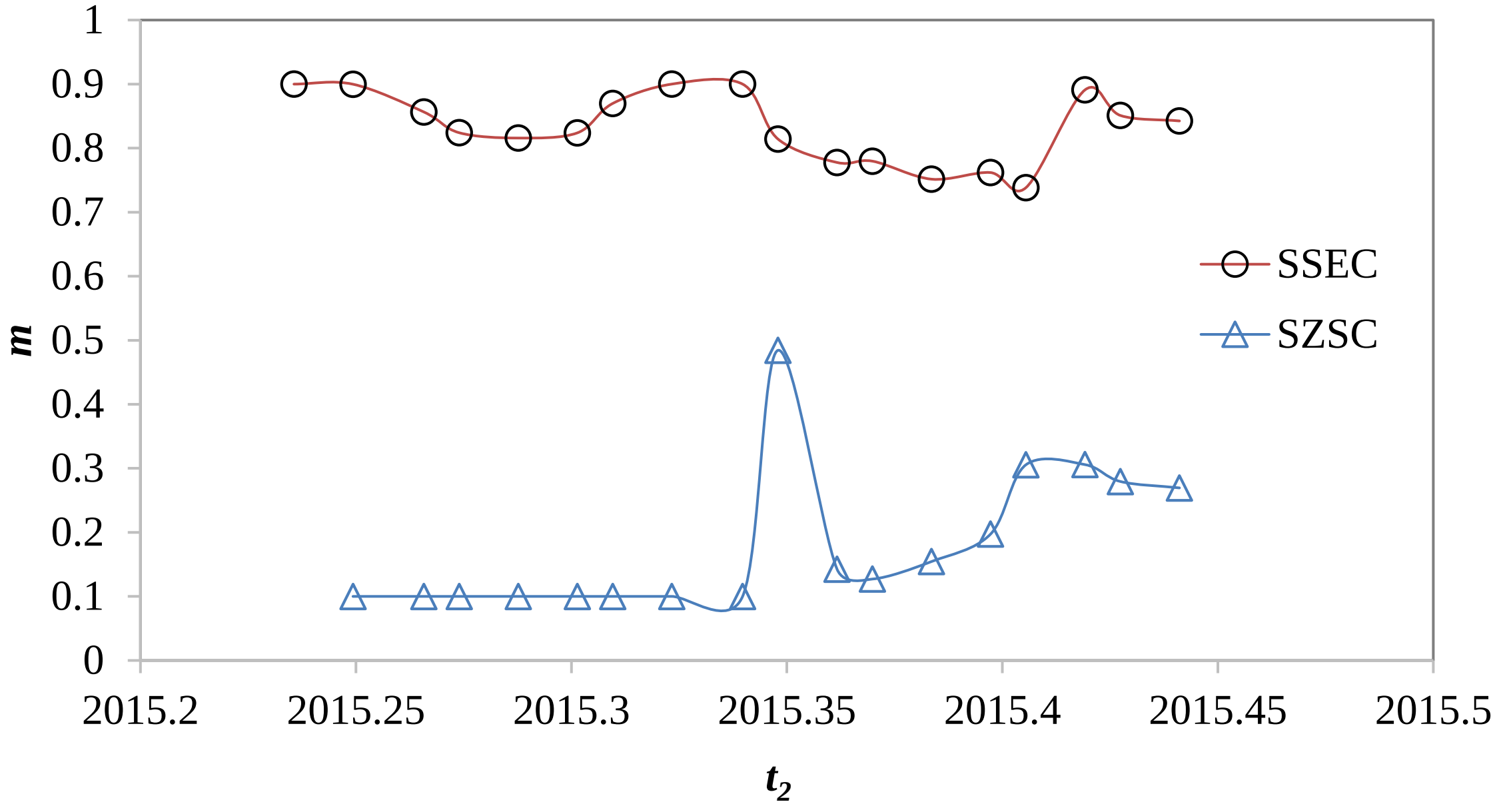

(a) The relationship between m and $t_2$

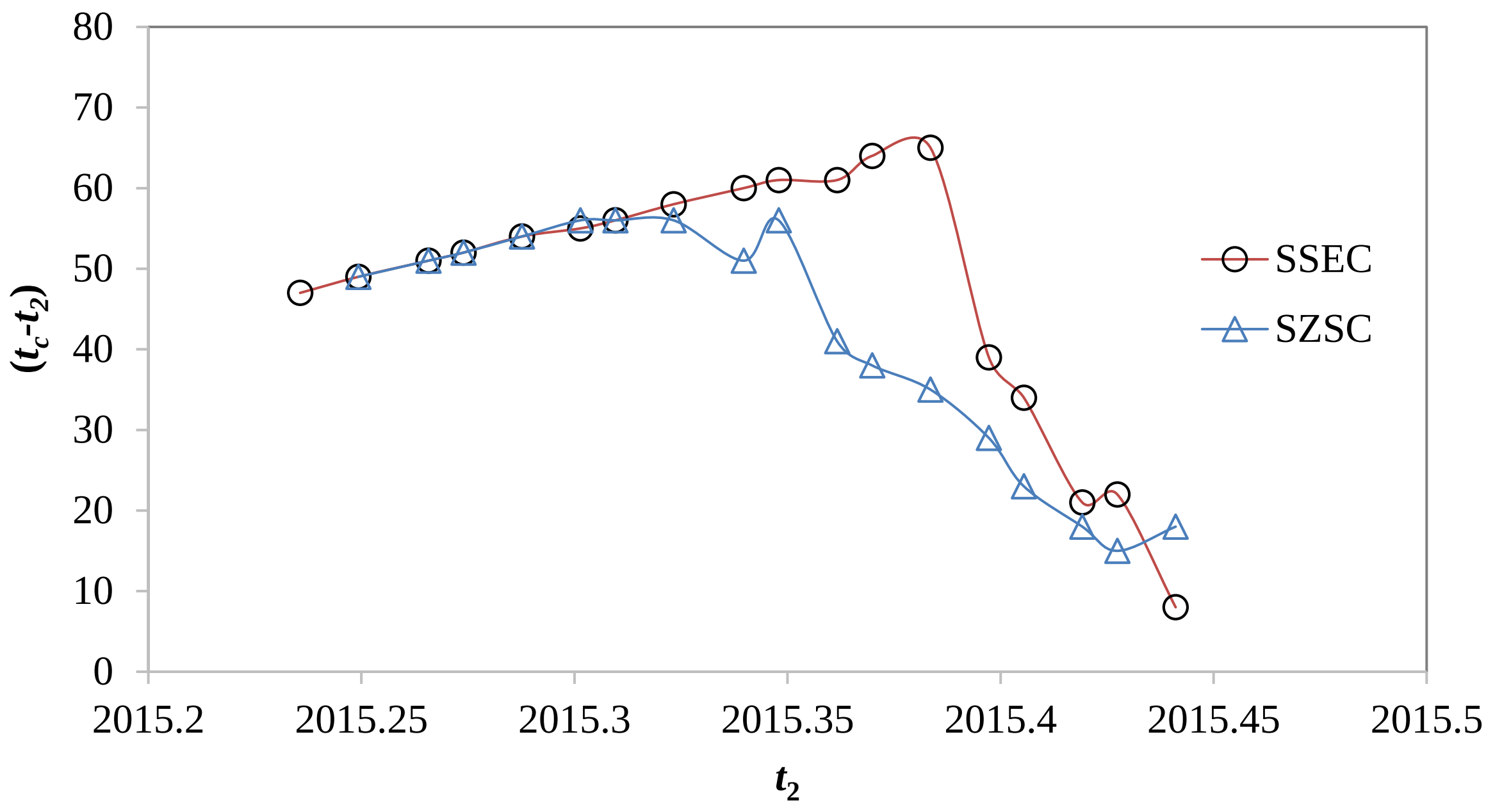

(b) The relationship between ( $t_c$ -$t_2$) and $t_2$

Figure 4. Change of the exponent m and the gap $(t_c - t_2)$ with the increasing end date $t_2$ in the expanding time interval with the fixed start date $t_1$=November 3, 2014 and the $t_2$ increasing from March 27, 2015 to June 10, 2015 in steps of three trading days.

*3.2 Lomb periodogram analysis*

The Lomb periodogram analysis is carried out to detect the logarithm-periodic oscillations in the LPPLS model for the 2015 Chinese Stock Market bubble. The results of the Lomb periodogram analysis on the detrended residual r(t) obtained from Equation (9) is summarized in the Figure 5. Figure 5 (a) presents the detrending residuals as a function of $\ln(t_c - t)$ using four typical examples, which are ( $t_1$ , $t_2$ )=(17 September 2014, 20 April 2015) and (3 November 2014, 7 May 2015) for SSEC and (17 Oct 2014, 20 April 2015) and (3 November 2014, 15 May 2015) for SZSC. The Lomb periodograms ($P_N$ with respect to $\omega_{\text{Lomb}}$) for the four examples are plotted in Figure 5 (b). The highest peak $P_N$ with its associate $\omega_{\text{Lomb}}$ are selected. The bivariate distribution of pairs ($\omega_{\text{Lomb}}$, $P_N^{Max}$) for different LPPLS calibration windows is presented in Figure 5 (c). Each point in Figure 5 (c) stands for the highest peak and the associated angular log-frequency in the Lomb periodogram for a given detrended residual series. For all the pairs ($\omega_{\text{Lomb}}$, $P_N^{Max}$) shown in Figure 5 (c), the false alarm probabilities are less than $10^{-5}$, indicating the true existence of the logarithm-periodic oscillations in the LPPLS model for the 2015 Chinese Stock Market bubble.

In general, the values of $\omega_{\text{Lomb}}$ is consistent with the values of $\omega_{\text{fit}}$ which is obtained from the LPPLS fitting procedures. Figure 5 (d) plots $\omega_{\text{fit}}$with respected to $\omega_{\text{Lomb}}$. It can be found that the most pairs of ($\omega_{\text{Lomb}}$, $\omega_{\text{fit}}$) are located around *y*=2*x*. It can be interpreted that the residuals have a fundamental log-periodic component at $\omega_{\text{Lomb}}$ and the harmonic component at $2\omega_{\text{Lomb}}$. The harmonic of log-periodic components can be expected to exist in log-periodic signals, which has been documented in early study of time series (Jiang et al., 2010; Sornette, 1998). It is in general a diagnostic of a significant log-periodic component when the harmonics have close-to-integer ratios to a common fundamental frequency. Since the value of $\omega_{\mathrm{fit}}$ in Equation (8) is bounded between 6 and 13 to ensure the log-periodic oscillations are neither too fast to fit a random component, nor too slow to support the trend, it is expected that there are no points around the line $y = x$ when the fundamental log-periodic components of residuals are not greater than the lower boundary of $\omega_{\mathrm{fit}}$ defined in Equation (8).

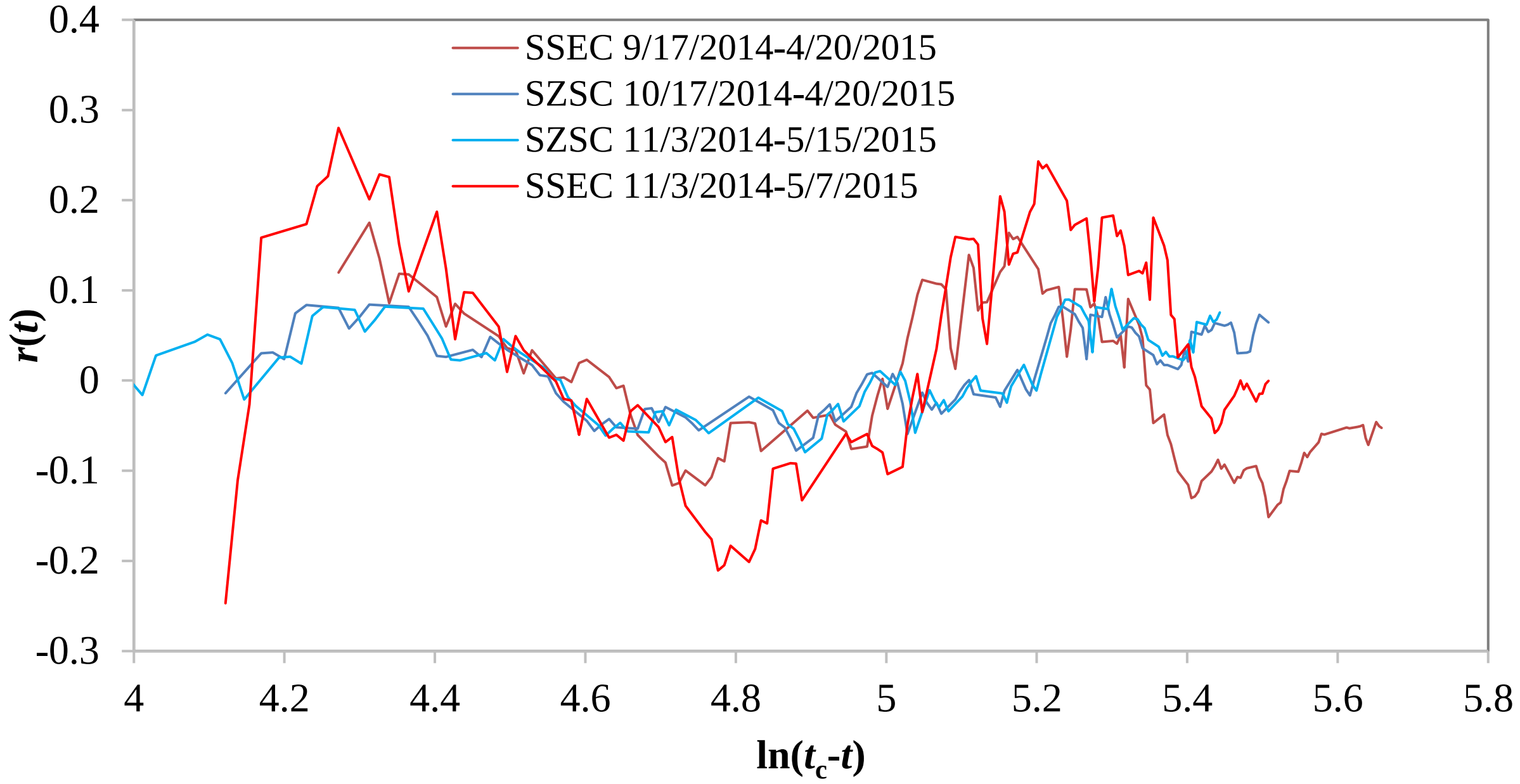

(a) Detrending residuals $r(t)$

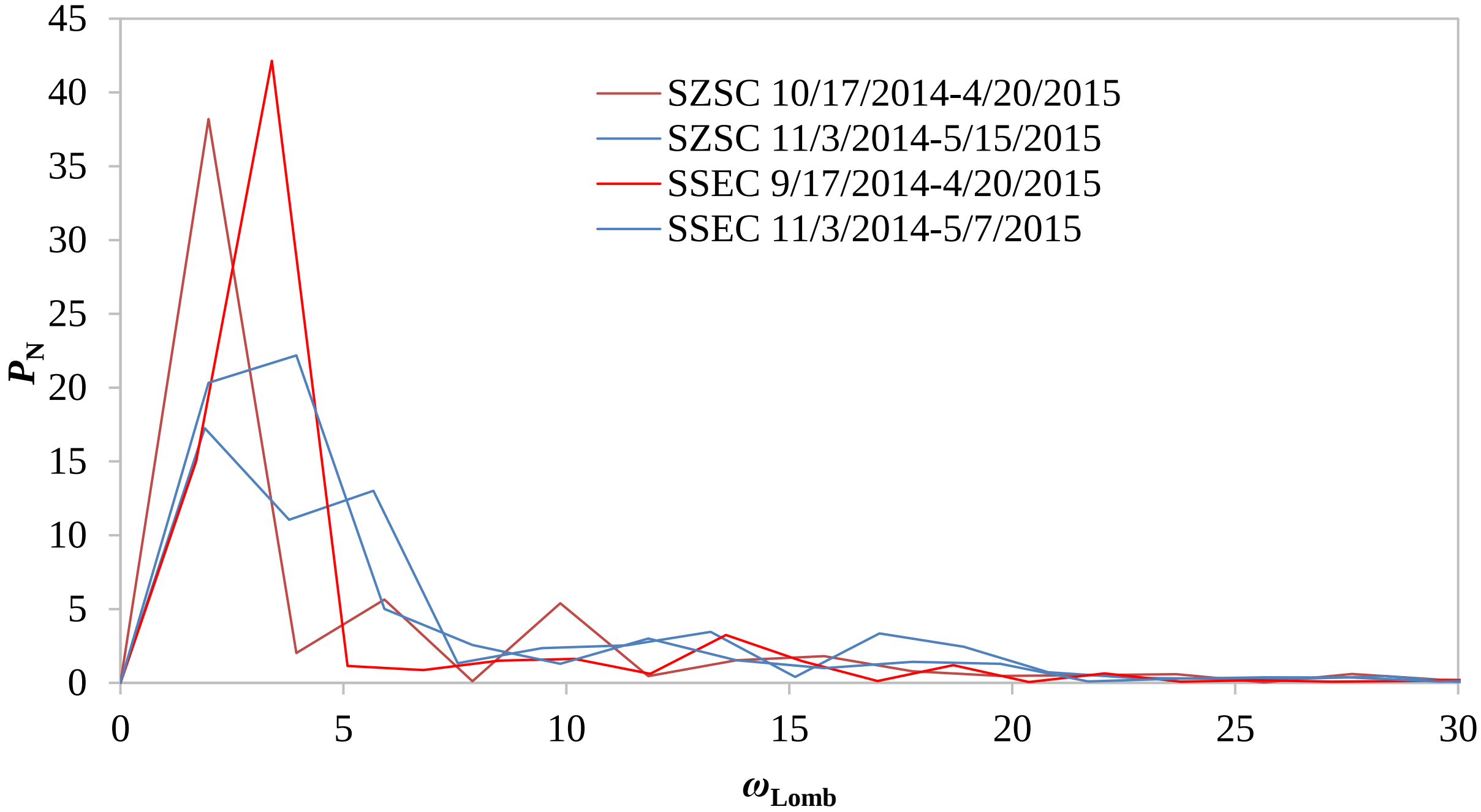

(b) Lomb periodograms for four typical examples.

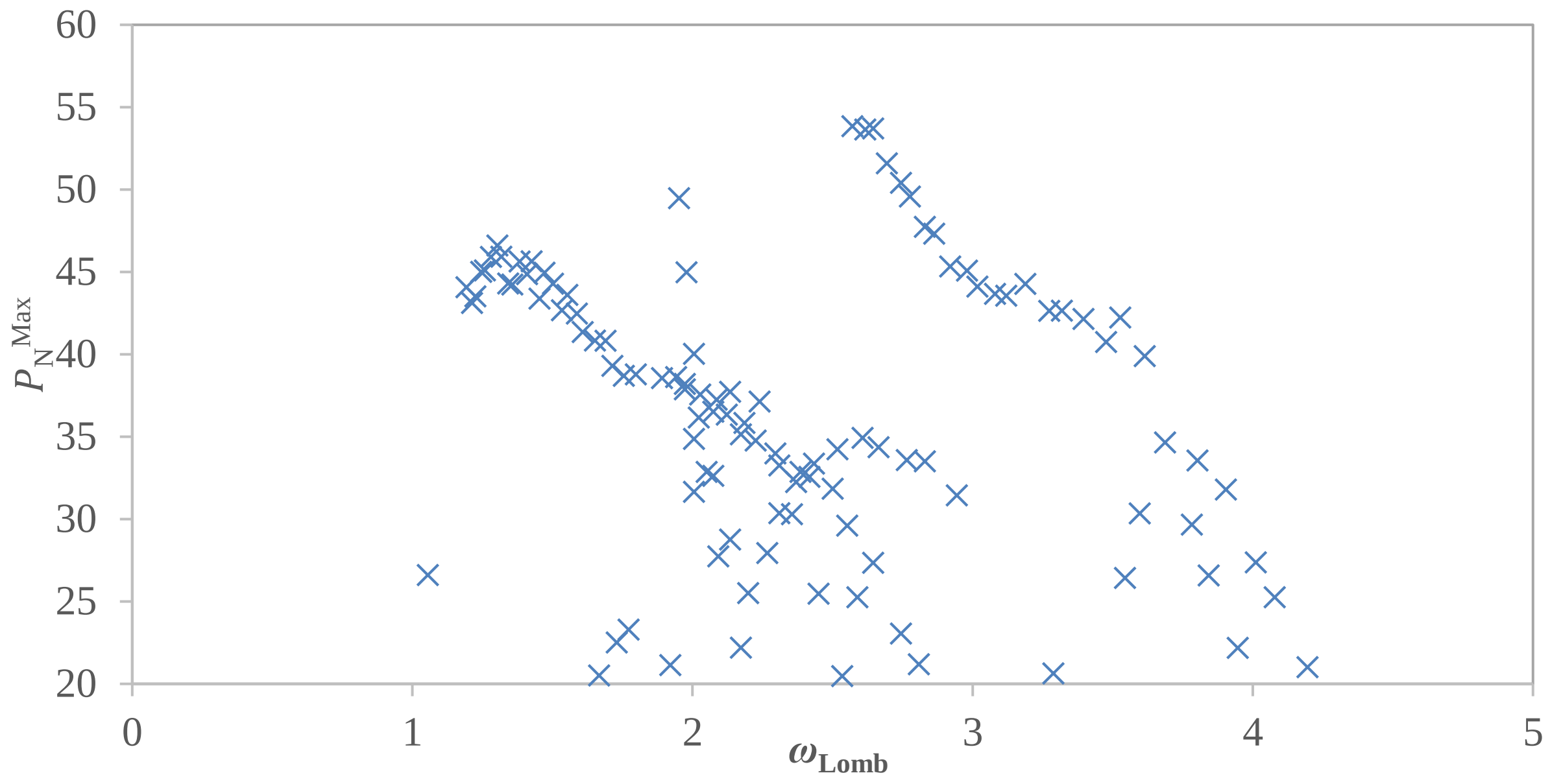

(c) Bivariate distribution of pairs ($\omega_{Lomb}$, $P_N^{Max}$)

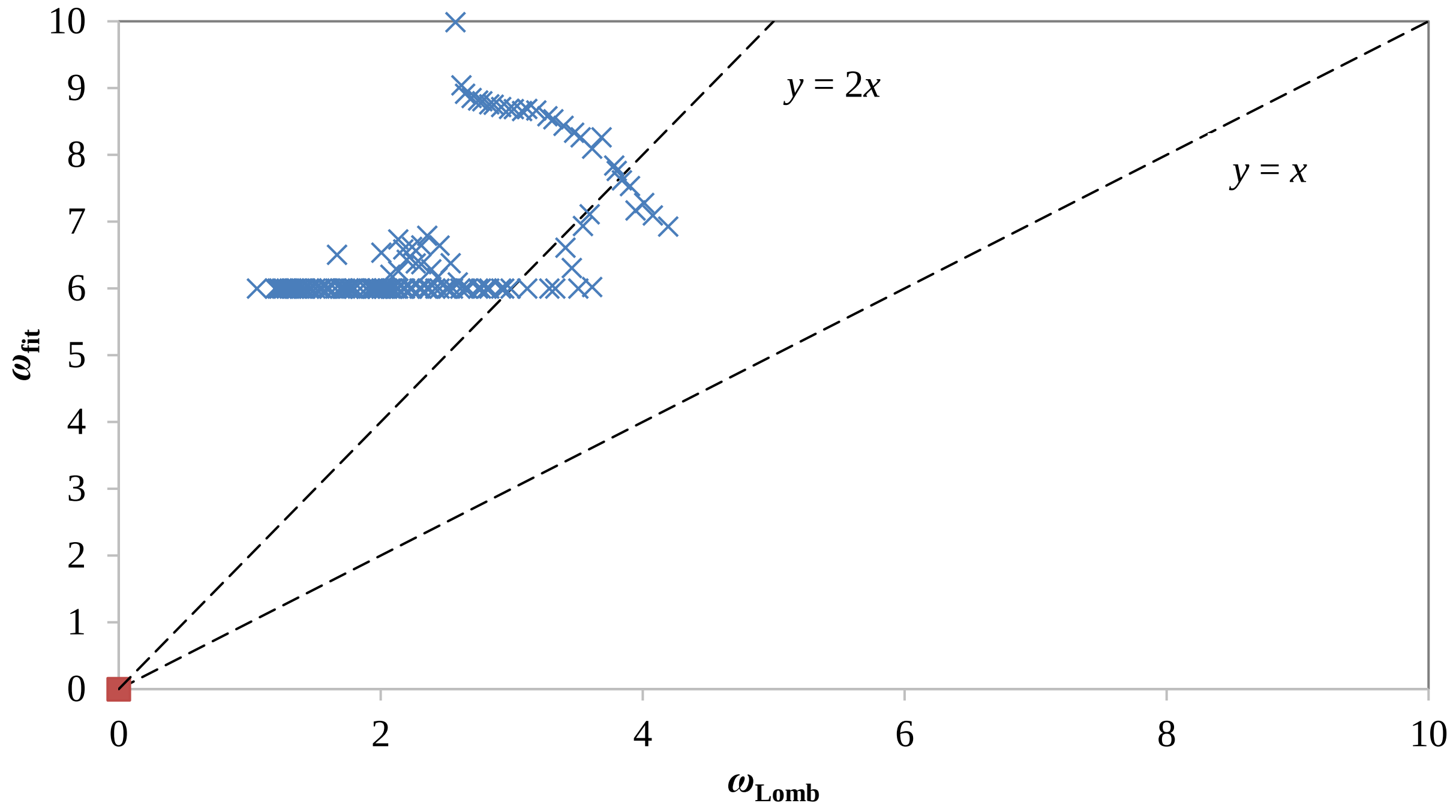

(d) The relationship between $\omega_{Lomb}$ and $\omega_{fit}$

Figure 5. Lomb tests of the detrending residuals r(t) for SSEC and SZSC

*3.3 Unit root tests of the 2015 Chinese stock market bubble*

To investigate the stationarity of the residuals between the logarithmic fitted price from the LPPLS model and the logarithmic observed price to determine if a mean-reversal Ornstein-Uhlenbeck (O-U) process can be applied to model the LPPLS fitting residuals, the unit root tests are used to the series of residuals for each $[t_1, t_2]$ interval. The null hypothesis $H_o$ of the unit root test is that the residuals are non-stationary. The residual time series has a unit root and are indeed stationary if the null hypothesis is rejected. Both the expanding and shrinking windows of the SSEC are scanned for each time interval. Results of these tests are listed in Table 1.

Table 1. Unit-root tests on the LPPLS fitting residuals for SSEC and SZSC index in the two window ranges. $N$ denotes the number of windows and $\alpha$ denotes the significant levels.

| Index | Windows range | $N$ | $\alpha$ | Percentage of rejecting H0 | |
|---|---|---|---|---|---|
| | | | | Phillips-Perron | Dickery-Fuller |
| SSEC | 2014/11/3-2015/6/10 | 16 | 0.05 | 100% | 100% |
| | | | 0.01 | 100% | 100% |
| SSEC | 2014/1/2-2015/4/20 | 79 | 0.05 | 96% | 100% |
| | | | 0.01 | 92% | 97% |
| SZSC | 2014/11/3-2015/6/10 | 17 | 0.05 | 100% | 100% |
| | | | 0.01 | 88% | 94% |
| SZSC | 2014/1/2-2015/4/20 | 75 | 0.05 | 95% | 97% |
| | | | 0.01 | 87% | 93% |

Table 1 shows that the minimum percentage of rejecting the null hypothesis $H_o$ at the significant level 0.05 based on the two tests is 95%, indicating the LPPLS fitting residuals are stationary. At the significant level of 0.01, the minimum percentage of rejecting $H_o$ is 87%. It can be interpreted as that the constraints in Equation (8) may not be enough to avoid the unreasonable fitting outcomes. Additional LPPLS filters are needed to improve the fitting performance of the LPPLS model.

## 4. Conclusions

In this study, we present a novel analysis of the 2015 financial bubble in the Chinese stock market by calibrating the LPPLS model to two important Chinese stock indices, SSEC and SZSC, from early 2014 to June 2015. The covariance matrix adaptation evolution strategy (CMA-ES) is adopted to search for the best estimators of the three nonlinear parameters $(t_c, m, \omega)$ to minimize the residuals (the sum of the squares of the differences) between the fitted LPPLS model and the observed price time series. The back tests indicate that the LPPLS model can well identify the bubble behavior of the faster-than-exponential increase corrected by the accelerating logarithm-periodic oscillations in the 2015 Chinese Stock market using both the SSEC and SZSC indices. The existence of log-periodicity is detected by applying the Lomb spectral analysis on the detrended residuals. The O-U property and stationarity in the residuals are confirmed by the two Unit-root tests (Philips-Perron test and Dickery-Fuller test) on the LPPLS fitting residuals.

While the post-mortem analysis of the 2015 Chinese stock market bubble is investigated in this study, it is emphasized to identify the bubbles and predict the critical time in advance of the demise of the bubble. According to our analysis, the LPPLS model may foretell the actual critical day two months before the actual bubble crash.

One challenge in implementing the LPPLS fitting procedure is the selection of multiple dimensional nonlinear optimization algorithm. Higher performance in the accuracy and lower computation cost are always our pursuit of the nonlinear optimization algorithm. Compared with the traditional optimization method used in LPPLS model, such as the taboo search and genetic algorithm, the CMA-ES may have a significantly low computation cost. It is recommended that the CMA-ES be used as a better algorithm for the LPPLS model fit. In order to expedite the fitting process, parallel computing is suggested to reduce the computation time drastically.

Moreover, the exponent $m$ which captures the mechanism of positive feedback leading to faster-than-exponential price growth in the LPPLS model, does not show a remarkable feature of change when the start day $t_1$ is fix and the end day $t_2$ is moved toward the actual critical time in the expanding windows. In the LPPLS fitting with expanding windows, the gap ($t_c$-$t_2$) shows a significant decrease when the end day $t_2$ is moved closer to the actual bubble crash time. The change rate of the gap ($t_c$-$t_2$) may be used as an additional indicator besides the key indicator $t_c$ to improve the accuracy of bubble burst prediction. Due to the limitation of boundary in the LPPLS fitting procedure, the fitted angular log-frequency $\omega_{fit}$ may have close-to-integer ratios to a common fundamental frequency of the detrended residual $r(t)$. Additional LPPLS filters may be needed to exclude the unreasonable fitting outcomes. The change rate of residuals in the LPPLS model is recommend as the additional filter to bound the search space in the fitting procedure.

## References

Auger, A., Hansen, N., 2005. A restart CMA evolution strategy with increasing population size, 2005 IEEE congress on evolutionary computation. IEEE, Edinburgh, UK, pp. 1769-1776.

Blanchard, O.J., Watson, M.W., 1982. Bubbles, rational expectations and financial markets, In: Wachtel, P. (Ed.), Crises in the Economic and Financial Structure, Lexington, MA, pp. 295-315.

Bothmer, H.-C., Meister, C., 2003. Predicting critical crashes? A new restriction for the free variables. Physica A 320, 539-547.

Brée, D.S., Challet, D., Peirano, P.P., 2013. Prediction accuracy and sloppiness of log-periodic functions. Quant. Finance 13, 275-280.

Demos, G., Sornette, D., 2017. Birth or burst of financial bubbles: which one is easier to diagnose? Quant. Finance 17, 657-675.

Filimonov, V., Demos, G., Sornette, D., 2017. Modified profile likelihood inference and interval forecast of the burst of financial bubbles. Quant. Finance 17, 1167-1186.

Filimonov, V., Sornette, D., 2013. A stable and robust calibration scheme of the log-periodic power law model. Physica A 392, 3698-3707.

Galbraith, J.K., 2009. The great crash 1929. Houghton Mifflin Harcourt, Boston, MA.

Gurkaynak, R., 2008. Econometric tests of asset price bubbles: Taking stock. J. Econ. Surv. 22, 166-186.

Hansen, N., Müller, S.D., Koumoutsakos, P., 2003. Reducing the time complexity of the derandomized evolution strategy with covariance matrix adaptation (CMA-ES). Evol. Comput. 11, 1-18.

Hansen, N., Ostermeier, A., 2001. Completely derandomized self-adaptation in evolution strategies. Evol. Comput. 9, 159-195.

Hansen, N., Ostermeier, A., Gawelczyk, A., 1995. On the Adaptation of Arbitrary Normal Mutation Distributions in Evolution Strategies: The Generating Set Adaptation, In: Eshelman, L. (Ed.), the Sixth International Conference on Genetic Algorithms. Morgan Kaufmann, San Francisco, CA, pp. 57-64.

Huang, Y., Johansen, A., Lee, M., Saleur, H., Sornette, D., 2000. Artifactual log-periodicity in finite size data: Relevance for earthquake aftershocks. J Geophys Res Solid Earth 105, 25451-25471.

Ide, K., Sornette, D., 2002. Oscillatory finite-time singularities in finance, population and rupture. Physica A 307, 63-106.

Igel, C., Hansen, N., Roth, S., 2007. Covariance matrix adaptation for multi-objective optimization. Evol. Comput. 15, 1-28.

Jacobsson, E., 2009. How to predict crashes in financial markets with the Log-Periodic Power Law, Department of Mathematical Statistics. Stockholm University.

Jiang, Z.-Q., Zhou, W.-X., Sornette, D., Woodard, R., Bastiaensen, K., Cauwels, P., 2010. Bubble diagnosis and prediction of the 2005–2007 and 2008–2009 Chinese stock market bubbles. J. Econ. Behav. Organ. 74, 149-162.

Johansen, A., Ledoit, O., Sornette, D., 2000. Crashes as critical points. Int. J. Theor. Appl. Finance 3, 219-255.

Johansen, A., Sornette, D., 1999. Financial" anti-bubbles": Log-periodicity in gold and Nikkei collapses. Int. J. Mod. Phys. C 10, 563-575.

Li, C., 2017. Log-periodic view on critical dates of the Chinese stock market bubbles. Physica A 465, 305-311.

Lin, L., Ren, R.E., Sornette, D., 2014. The volatility-confined LPPL model: A consistent model of ‘explosive’ financial bubbles with mean-reverting residuals. Int. Rev. Finan. Anal. 33, 210-225.

Lux, T., Sornette, D., 2002. On rational bubbles and fat tails. J. Money Credit Bank. 34, 589-610.

Shiller, R.J., 2015. Irrational exuberance. Princeton University Press, Princeton, NJ.

Sornette, D., 1998. Discrete-scale invariance and complex dimensions. Phys. Rep. 297, 239-270.

Sornette, D., 2003. Critical market crashes. Phys. Rep. 378, 1-98.

Sornette, D., Demos, G., Zhang, Q., Cauwels, P., Filimonov, V., Zhang, Q., 2015. Real-time prediction and post-mortem analysis of the Shanghai 2015 stock market bubble and crash. J. Invest. Strategies 4, 77–95.

Sornette, D., Johansen, A., 1997. Large financial crashes. Physica A 245, 411-422.

Sornette, D., Woodard, R., Zhou, W.-X., 2009. The 2006–2008 oil bubble: Evidence of speculation, and prediction. Physica A 388, 1571-1576.

Sornette, D., Zhou, W.X., 2002. The US 2000-2002 market descent: How much longer and deeper? Quant. Finance 2, 468-481.

Stiglitz, J.E., 2014. The lessons of the North Atlantic crisis for economic theory and policy. Cambridge, MA: MIT Press.

Yan, W., 2011. Identification and Forecasts of Financial Bubbles, Department of Management, Technology and Economics ETH Zurich.

Yan, W., Woodard, R., Sornette, D., 2012. Diagnosis and prediction of market rebounds in financial markets. Physica A 391, 1361-1380.
Zhang, Q., Sornette, D., Balcilar, M., Gupta, R., Ozdemir, Z.A., Yetkiner, H., 2016. LPPLS bubble indicators over two centuries of the S&P 500 index. Physica A 458, 126-139.
Zhou, W., Sornette, D., 2003. 2000–2003 real estate bubble in the UK but not in the USA. Physica A 329, 249-263.
Zhou, W., Sornette, D., 2006. Is there a real-estate bubble in the US? Physica A 361, 297-308.
Zhou, W., Sornette, D., 2008. Analysis of the real estate market in Las Vegas: Bubble, seasonal patterns, and prediction of the CSW indices. Physica A 387, 243-260.